\documentclass[fleqn,usenatbib]{mnras}
\usepackage[T1]{fontenc}
\usepackage{ae,aecompl}
\usepackage{bm}
\usepackage{graphicx}

\usepackage{amsmath}	
\usepackage{amssymb}	
\usepackage{txfonts}

\title{Can a binary star host three giant circumbinary planets?}

\author[Chen et al.]{Cheng Chen$^{1}$\thanks{Email: c.chen6@leeds.ac.uk}, Rebecca G. Martin$^{2,3}$ and C. J. Nixon$^1$ 
\\ $^{1}$School of Physics and Astronomy, University of Leeds, Leeds LS2 9JT, UK
\\ $^2$Department of Physics and Astronomy,  University of Nevada, Las Vegas, 4505 South Maryland Parkway, Las Vegas, NV 89154, USA 
\\ $^{3}$Nevada Center for Astrophysics, University of Nevada, Las Vegas,
4505 South Maryland Parkway, Las Vegas, NV 89154, USA\\
}

\date{Accepted XXX. Received YYY; in original form ZZZ}

\pubyear{2023}

\begin{document}
\label{firstpage}
\pagerange{\pageref{firstpage}--\pageref{lastpage}} 
\maketitle


\begin{abstract}
We investigate the orbital stability of a tilted circumbinary planetary system with three giant planets. The planets are spaced by a constant number ($\Delta$) of mutual Hill radii  in the range $\Delta=3.4-12.0$ such that the period ratio of the inner pair is the same as the outer pair. A tilted circumbinary planetary system can be unstable even if the same system around a coplanar binary is stable. For an equal mass binary, we find that the stability of a three-planet system is qualitatively similar to that of a two-planet system, but the three-planet system is more unstable in mean motion resonance regions. For an unequal mass binary, there is significantly more instability in the three-planet system as the inner planets can undergo von-Zeipel-Kozai-Lidov oscillations. Generally in unstable systems, the inner planets are more likely to be ejected than the outer planets. The most likely unstable outcome for closely spaced systems, with $\Delta \lesssim 8$, is a single remaining stable planet. For more widely separated systems, $\Delta \gtrsim 8$, the most likely unstable outcome is two stable planets, only one being ejected. An observed circumbinary planet with  significant eccentricity may suggest that it was formed from an unstable system. Consequently, a binary can host three tilted giant planets if the binary stars are close to equal mass and provided that the planets are well spaced and not close to a mean motion resonance.


\end{abstract}


\section{Introduction}

Although the multiplicity distribution of circumbinary planets (CBPs) has not been confirmed, about half of planets around single stars are known to have siblings in the Kepler data \citep[e.g.][]{Berger2018, Thompson2018, Sandford2019}.  In the radial velocity planet sample of the California Legacy Survey, 30 -- 40$\%$ of Sun-like stars with a massive planet (mass $>0.3\,M_{\rm J}$, where $M_{\rm J}$ is the mass of Jupiter) are multi-planet systems  \citep{ZW2022}. Currently, there are only five multi-planet systems that have been found around binaries and all of the CBPs in these systems are nearly coplanar to the binary orbital plane. In the Kepler-47 system, there are three Neptune-size planets with circular ($e_{\rm p} < 0.03$) orbits  \citep{Orosz2012a, Orosz2012b, Kostov2013}. In the TOI-1338 system, there are two saturnian CBPs found by the transit and radial velocity methods \citep{Kostov2020, Standing2023a}. The red dwarf -- white dwarf binary system, NN Ser, has two Jupiter-mass CBPs \citep{Mustill2013}. Kepler-451 hosts three Jupiter-mass CBPs  \citep{Baran2015, Esmer2022}. However, the coplanarity of these systems is likely a result of observational bias \citep[e.g.][]{MartinTriaud2015,MartinD2017,MartinD2017b}. So, while it has been established from observations that it is possible to generate circumbinary planetary systems with 3 or more giant planets in the coplanar case, we do not yet know anything about the non-coplanar case, or the more general stability properties of the coplanar case.

Despite the fact that only coplanar CBPs have currently been detected, recent observations have found that there are many misaligned circumbinary discs \citep[e.g.,][]{Chiang2004,Winn2004,Capelo2012,Kennedy2012,Brinch2016,Kennedy2019,Zhu2022,Kenworthy2022}. Polar aligned discs around eccentric binaries may also be common \citep{Aly2015,Kennedy2019,Smallwood2020,Kenworthy2022}. Circumbinary planets  may form inside these misaligned discs.  Theoretically, the formation of misaligned circumbinary discs can result from chaotic accretion \citep{clarke1993,Bate2018} and subsequent disc evolution can lead to tilt evolution towards a coplanar alignment \citep{Bateetal2000,Lubow2000,Nixonetal2011a} or a polar alignment \citep{Martin2017,MartinandLubow2018b,Lubow2018,Zanazzi2018, Abod2022}. Although about 68\% of short period binaries (period $<20\,\rm days $) have aligned disks (within 3$^{\circ}$), those with longer orbital periods have a larger range of inclinations and binary eccentricities \citep{Czekala2019}. If the alignment timescale for an extended disc is longer than the disc lifetime then planetary systems may form in a misaligned disc. Misaligned planetary systems may be expected around binaries with a longer orbital period \citep{Czekala2019,MartinandLubow2019}. These planetary systems may be observed in the future with eclipse-timing variations \citep{Zhang2019,MartinD2019}.

A misaligned CBP has a complicated interaction with the binary. For any inclination around a circular orbit binary, or for low inclination around an eccentric orbit binary, the angular momentum vector of a misaligned CBP  precesses around the binary angular momentum vector. The longitude of the ascending node fully circulates over 360$^\circ$ during the nodal precession, these are called {\it circulating} orbits. Around an eccentric orbit binary, the angular momentum vector a misaligned CBP orbit  may precess about the binary eccentricity vector. These are called {\it librating} orbits \citep{Verrier2009,Farago2010,Doolin2011,Naoz2017,deelia2019,Chen20192}.  The minimum inclination (critical inclination) for libration decreases with increasing binary eccentricity ($e_{\rm b}$). Therefore, a CBP orbits with even a small initial inclination can librate around a highly eccentric binary. During nodal libration, a CBP undergoes tilt oscillations while its longitude of  ascending node is limited in a range of angles less than 360$^\circ$.

Close to a binary, even a single planet system can be unstable \citep[e.g.][]{Holman1997}. Both nodal precession of the orbit and mean motion resonances (MMRs) with the binary play roles in the stability  \citep{Doolin2011, Sutherland2016}. In general, a CBP with a circular orbit is stable if its semi-major axis is greater than about 5  times separation of the binary, $a_{\rm b}$ for planet masses up to about $m_{\rm p}=10\,M_{\rm J}$ \citep{Doolin2011, Chen20201}. However,  stable orbits can exist  closer in, down to around 2$a_{\rm b}$, when the CBP is in a nearly retrograde orbit for small binary eccentricity \citep{Hong2019,Cuello2019, Giuppone2019} or in a polar orbit for high $e_{\rm b}$ \citep{Chen20201}.

For a circumbinary planetary system with two massive planets, the dynamics are more complicated because of planet-planet interactions \citep{Chen2022}. These  lead to tilt oscillations between the planets, MMRs between the planets and von-Zeipel-Kozai-Lidov \citep[ZKL][]{vonZeipel1910,Kozai1962,Lidov1962} oscillations of the inner planet that lead to planet eccentricity growth \citep{Chen2023}.  The combination of planet-planet and planet-binary interactions can result in a planet being ejected from the system, as several simulations have already shown in coplanar CBP systems \citep[e.g.,][]{Smullen2016, Sutherland2016, Gong20171, Gong20172}. The tilt of a circumbinary planetary system with respect to the binary orbit has a large influence in the stability of the system. Planetary systems which are stable around a coplanar binary become unstable for  a wide range of parameters in a tilted system \citep{Chen2023}.

On the other hand, around a single star, planet-planet scattering occurs only when the planets form very close to each other. Two planets with masses $m_{\rm p1}$ and $m_{\rm p2}$ that form with semi-major axes $a_{\rm p1}$ and $a_{\rm p2}$, respectively, around a star with mass $m_{\rm b}$ are unstable if  $\Delta\lesssim 2\sqrt{3}$, where we define
\begin{equation}
\Delta=\frac{a_{\rm p2}-a_{\rm p1}}{R_{\rm Hill}},
\end{equation}
and the mutual Hill radius between two planets, $i$ and $i+1$, is 
\begin{equation}
R_{\rm Hill}=\left(\frac{m_{\rm pi}+m_{{\rm p}(i+1)}}{3 \,m_{\rm b}}\right)^{1/3}\left(\frac{a_{{\rm p}i}+a_{{\rm p}(i+1)}}{2} \right),
\end{equation}
\citep{Marchal1982,Gladman1993,Chambers1996}. However, in observed planetary systems,  93$\%$ of planet pairs are greater than 10$\, R_{\rm Hill}$ apart and 20$\,R_{\rm Hill}$ apart is the most common separation in the California--Kepler Survey \citep{Weiss2018}. 

This stability criterion is not significantly affected if a single star is replaced by a {\it coplanar} inner binary, unless the planets are formed very close to the binary \citep{Kratter2014}. However, the outcome of an unstable system is more likely to be ejection rather than collision around a binary star \citep{Smullen2016, Sutherland2016,Gong20171,Gong20172,Fleming2018}. This is because of close encounters with the binary. Coplanar planets around a binary must form close to each other to be unstable.


The  instability and ejection of CBPs may be a mechanism to produce  free-floating planets (FFPs). Gravitational microlensing observations suggest that there are more FFPs than main-sequence stars by a factor of $1-3.5$ \citep[e.g.][]{Sumi2011}. There is an excess of planets by a factor of up to seven compared to that predicted by core-collapse models \citep{Padoan2002,Miret-Roig2021}. There are several mechanisms suggested to form the excess of free-floating planets. These include planet-planet scattering \citep{Rasio1996,Weidenschilling1996,Veras2012}, aborted stellar embryo ejection from a stellar nursery \citep{Reipurth2001} and photo-erosion of a pre-stellar core  by stellar winds from a nearly OB star \citep{Whitworth2004}. Planet-planet scattering around single stars is one possible mechanism but it cannot explain the large number of free-floating planets \citep{Veras2012}.


In this paper, we model circumbinary systems with three Jupiter-mass CBPs and examine their stability. We consider how many planets can be ejected from a system and the masses of the ejected planets. In Section~\ref{five} we first describe the setup of our simulations and then we show  stability maps in Section~\ref{sim} and the final distribution of surviving planets in Section~\ref{dis}.  Finally, we address our discussion and conclusions in Section~\ref{con} and implications for observations of circumbinary planets in Section~\ref{implications}.

\section{Five-body simulations}
\label{five} 

To study the orbital stability of three planet systems orbiting around a circular or eccentric binary star, we carry out  simulations with the ${\sc n}$-body simulation package, {\sc rebound} with a {\sc whfast} integrator which is a second order symplectic Wisdom Holman integrator with 11th order symplectic correctors \citep{Rein2015b}. We solve the gravitational equations for the five bodies in the
frame of the centre of mass of the five-body system. The central binary has components of mass $m_1$ and $m_2$ with a total mass of $m_{\rm b}=m_1+m_2$ and mass fraction of $f_{\rm b}=m_2/m_{\rm b}$. The binary is in a circular or eccentric orbit with the binary eccentricity $e_{\rm b}$  and their separation $a_{\rm b}$ = 1.0 in simulations.

\subsection{Simulation set--up and parameter space explored}
\label{sta}

The three planets have equal masses $m_{\rm p1} = m_{\rm p2} = m_{\rm p3} = 0.001\,m_{\rm b}$.  The planets are initially in circular Keplerian orbits around the centre of mass of the binary. Our simulations do not consider  collisions or the formation of S-type planets (planets which orbit around one star of a binary). The planet orbits are defined by six orbital elements: the semi-major axes $a_{\rm p1}$,  $a_{\rm p2}$ and $a_{\rm p3}$, inclinations relative to the binary orbital plane $i_{\rm p1}$, $i_{\rm p2}$ and  $i_{\rm p3}$, eccentricities $e_{\rm p1}$, $e_{\rm p2}$ and $e_{\rm p3}$, longitude of the ascending nodes measured from the binary semi--major axis $\phi_{\rm p1}$, $\phi_{\rm p2}$ and $\phi_{\rm p3}$, argument of periapsides $\omega_{\rm p1}$, $\omega_{\rm p2}$ and $\omega_{\rm p3}$, and true anomalies $\nu_{\rm p1}$, $\nu_{\rm p2}$ $\nu_{\rm p3}$. The initial orbits of the three planets are coplanar to each other and circular so initially $i_{\rm p1}=i_{\rm p2}=i_{\rm p3}$, $e_{\rm p}=0$, $\omega_{\rm p}=0$ and $\nu_{\rm p}=0$ and we set $\phi_{\rm p} =90^{\circ}$ for all planets.

To understand the dynamic interactions between three planets, the inner planet is placed at $a_{\rm p1}=5\,a_{\rm b}$, where a single CBP is stable  for all initial inclinations \citep{Chen20201}.
We calculate the semi-major axis of the second and third planets such that they are separated by a fixed number, $\Delta$, of mutual Hill radii.
Thus, the semi-major axes of the planets are related by
\begin{equation}
    a_{{\rm p}(i+1)} = a_{{\rm p}i} + \Delta \, R_{\rm Hill}
\end{equation}
for $i=1,2$. 
The planet masses are fixed and $a_{\rm p1}$ is chosen. For a fixed value for $\Delta$, this equation can first be solved to find $a_{\rm p2}$ and then with that solution we can then solve it again to find $a_{\rm p3}$. We consider $\Delta$ in the range from 3.4 $\left(\approx 2\sqrt{3} \right)$ to 12.0. 
We integrate the simulations for a total time of 14 million binary orbital periods ($T_{\rm b}$).
The timestep of integration is 0.7$\%$ of the initial orbital period of the inner planet. We list the parameters of the four models that we explore in Table~\ref{table1}. We define the orbit of the planet as unstable once at least one of three criteria are met.  Instability occurs first, if the eccentricity of the planet becomes large $e_{\rm p} \geq 1.0$ so that the planet is not bound to the binary; second, if the semi--major axis of the planet increases significantly,  $a_{\rm p} > 1000\, a_{\rm b}$; or third, if the semi-major axis of the planet is smaller than the binary separation, $a_{\rm p} <  a_{\rm b}$ \citep[see also, for example][]{Quarles2019}.  

\begin{table}
\centering
\caption{Parameters of the simulations. The first column contains the name of the Model, the second and third columns indicate the binary mass fraction and the binary eccentricity. The fourth and fifth columns are
minimum and maximum difference of the semi-major axis between two planets which we consider with an interval of $\Delta = 0.1$. }
\begin{tabular}{ccccccc} 
\hline
\textbf{Model} & $f_{\rm b}$ & $e_{\rm b}$ & $a_{\rm p1}$ ($a_{\rm b}$) & min. $\Delta $ & max. $\Delta$  \\
\hline
\hline

T1 & 0.5 & 0.0  & 5.0 & 3.4 & 12.0 \\
T2 & 0.5 & 0.8  & 5.0 & 3.4 & 12.0 \\
T3 & 0.1 & 0.0  & 5.0 & 3.4 & 12.0  \\
T4 & 0.1 & 0.8  & 5.0 & 3.4 & 12.0  \\

\hline
\label{table1}
\end{tabular}
\end{table}

\subsection{Stability maps: two circumbinary planets}

\begin{figure*}
  \centering
    \includegraphics[width=8.7cm]{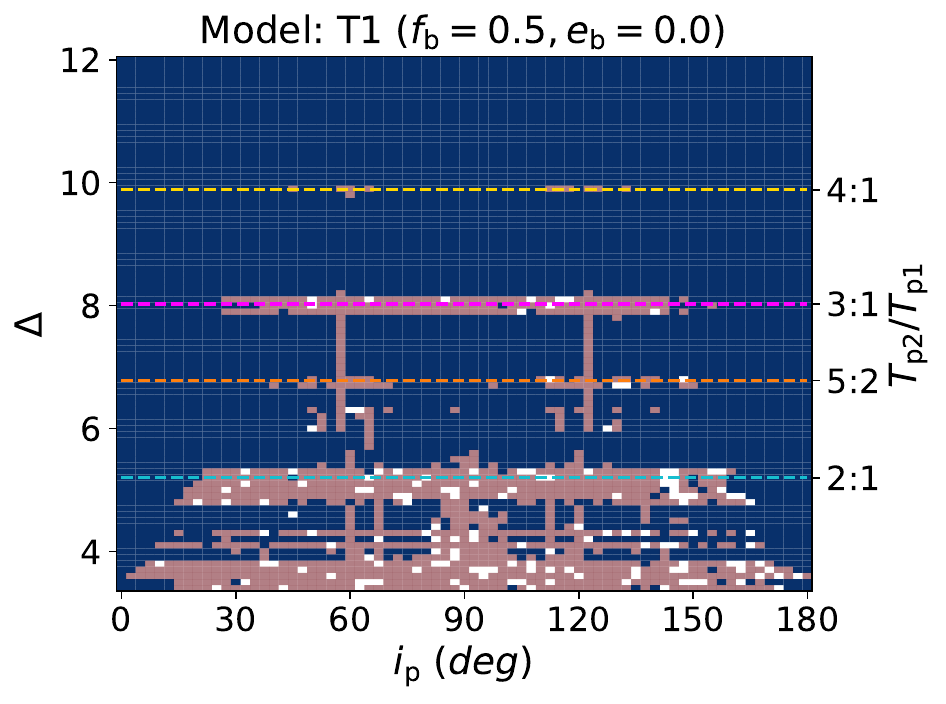}
    \includegraphics[width=8.7cm]{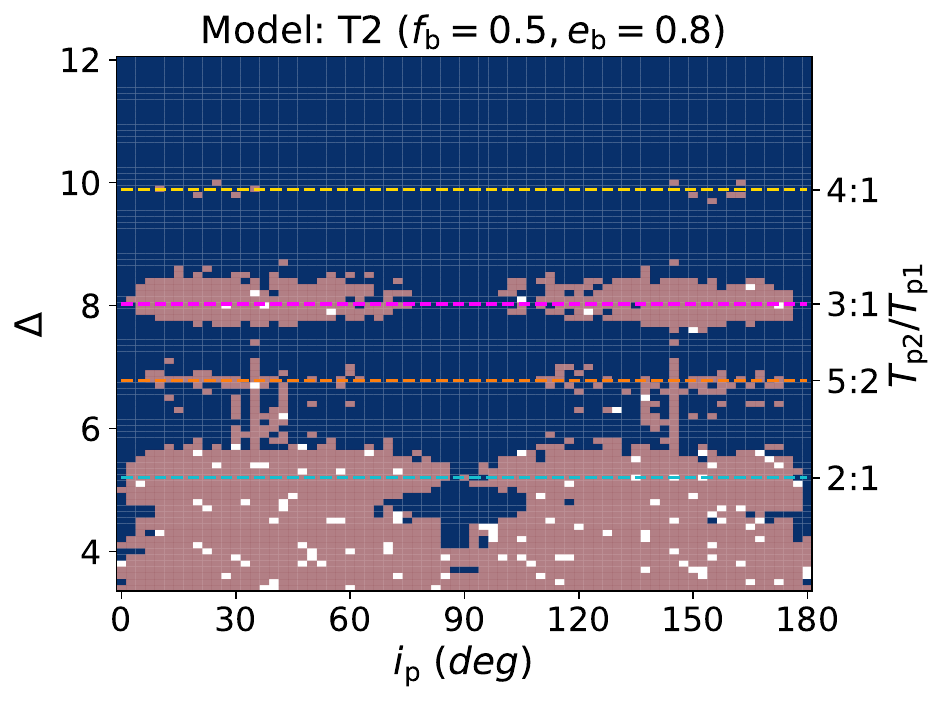}
    \includegraphics[width=8.7cm]{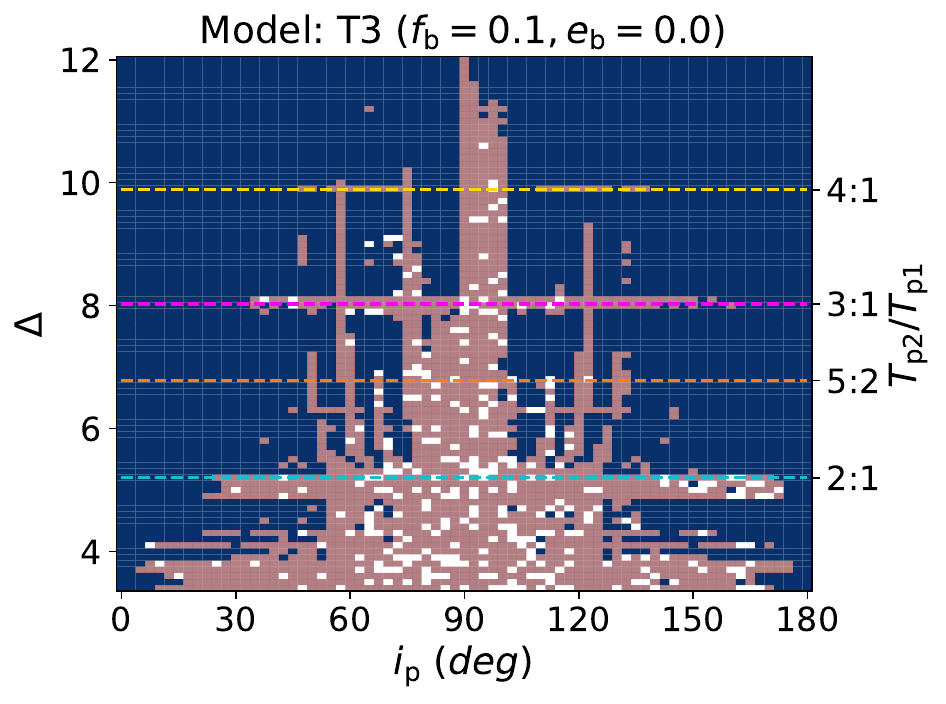}
    \includegraphics[width=8.7cm]{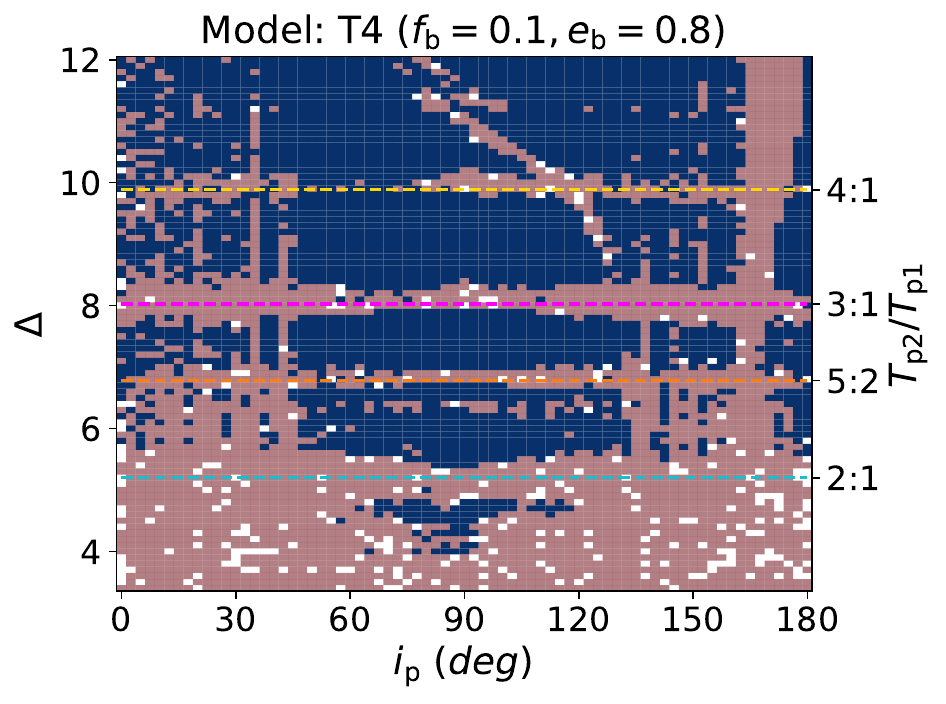}       
    \caption{Stability maps for planetary systems with two circumbinary planets around a binary with $e_{\rm b} =0.0$ and $f_{\rm b}=0.5$ (upper-left), $e_{\rm b} =0.8$ and $f_{\rm b}=0.5$ (upper-right panel), $e_{\rm b} =0.0$ and $f_{\rm b}=0.1$ (lower-left) and $e_{\rm b} =0.8$ and $f_{\rm b}=0.1$ (lower-right). The x-axis is the initial planet inclination $i_{\rm p}$ and the y-axis is the separations of the planets, $\Delta$, in units of their mutual Hill Radii. The inner planet has initial semi-major axis $a_{\rm p1} = 5\, a_{\rm b}$. The four horizontal dashed lines are the 2:1, 5:2, 3:1 and 4:1 MMR between the inner and middle planets.  Blue pixels represent systems in which two planets are stable, red pixels represent systems in which only one planet is stable and white pixels represent systems in which all the planets are unstable.}
     \label{fig:two}
\end{figure*}

To compare with the stability maps for three planet systems, we first run simulations of two planet systems by removing the outer planet from the simulation. 
Fig.~\ref{fig:two} shows stability maps  for varying planetary system separation, $\Delta$, and initial inclination, $i_{\rm p}$. The colours of the pixels indicate the stability of the system at the end of each simulation. Blue pixels represent systems in which two CBPs are stable, red pixels represent systems in which only one CBP survived and white pixels represent systems in which both CBPs are unstable at the end of the simulation. The four horizontal dashed lines indicate the 2:1, 5:2, 3:1 and 4:1 MMRs between the two CBPs.
The upper two panels are similar to the maps in  figure 2 of \citet{Chen2023} but extended to  larger planet separation. There are two unstable regions around the 3:1 MMR and the 4:1 MMR that were not previously seen. The unstable regions in model T2 (top right) are wider than the that of in model T1 (top left) due to the larger binary eccentricity $e_{\rm b}$.

The lower-left panel shows the stability map for a circular binary system with $f_{\rm b} = 0.1$. The system is quite unstable between $i_{\rm p} = 50^{\circ} \sim 130^{\circ}$ until the planets are more widely separated than their 2:1 MMR. There are several vertical unstable belts that are similar to the upper-left panels of figure 3 in \citet{Chen2023} in which the inner planet was located farther out,  at 10$a_{\rm b}$. This is because the nodal oscillation timescale is proportional to $1/(f_b (1-f_b))$ \citep{Lubow2018}. A smaller binary mass ratio leads to slower nodal precession. Therefore the planet-planet interactions become more important compared to the binary-planet interactions.  A smaller $f_{\rm b}$ is equivalent to placing the innermost planet at a larger distance to the binary.  As a result, the stability map with $f_{\rm b} = 0.1$ is similar to that of with $f_{\rm b}$ = 0.5 and two CBPs at larger $a_{\rm p}$.

The lower-right panel shows the stability map for a binary with $e_{\rm b}= 0.8$ and $f_{\rm b} = 0.1$. There are very few stable orbits around the polar region inside of the 2:1 MMR. Outside of  the 2:1 MMR, two CBPs are more stable in the region between $i_{\rm p} = 50^{\circ} \sim 130^{\circ}$ except, around MMR regions, compared to the coplanar and retrograde regions. Two CBPs are likely to be unstable even in coplanar orbits while CBPs in retrograde orbits are relatively stable although there is a vertical unstable belt around  $i_{\rm p} = 170^{\circ}$. Further, the region around the 4:1 MMR is more unstable compared to the other three maps.

\subsection{Stability maps: three circumbinary planets}
\label{sim}

\begin{figure*}
  \centering
    \includegraphics[width=8.7cm]{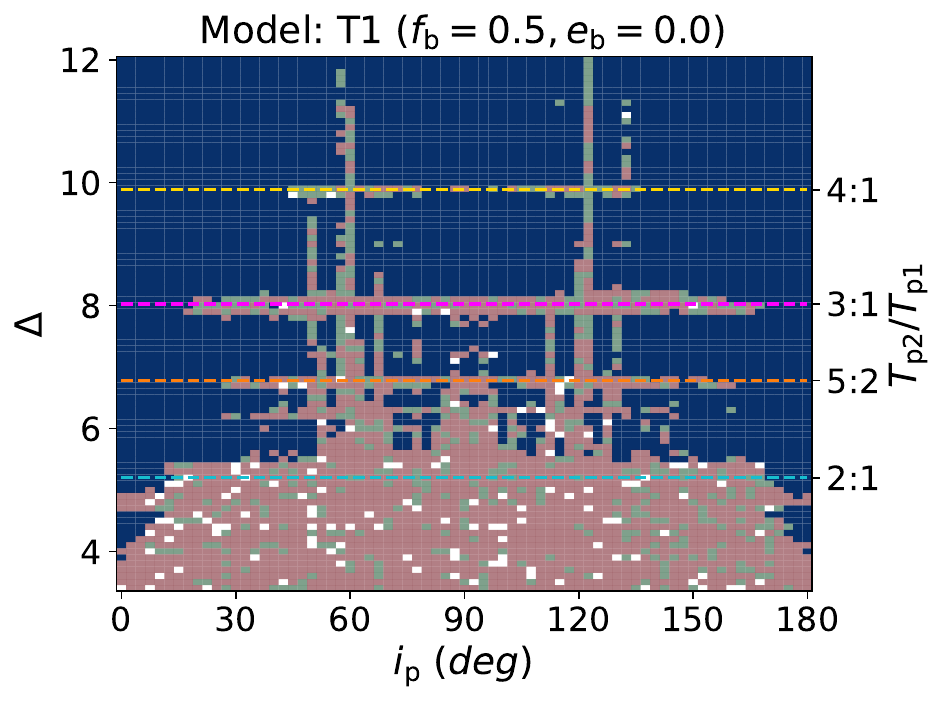}
    \includegraphics[width=8.7cm]{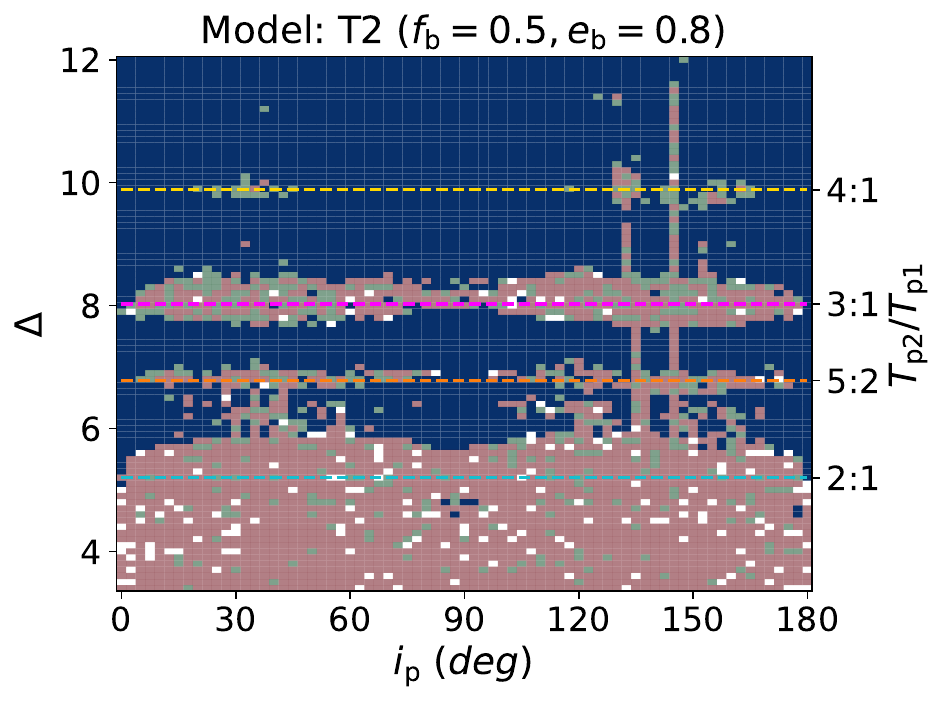}
    \includegraphics[width=8.7cm]{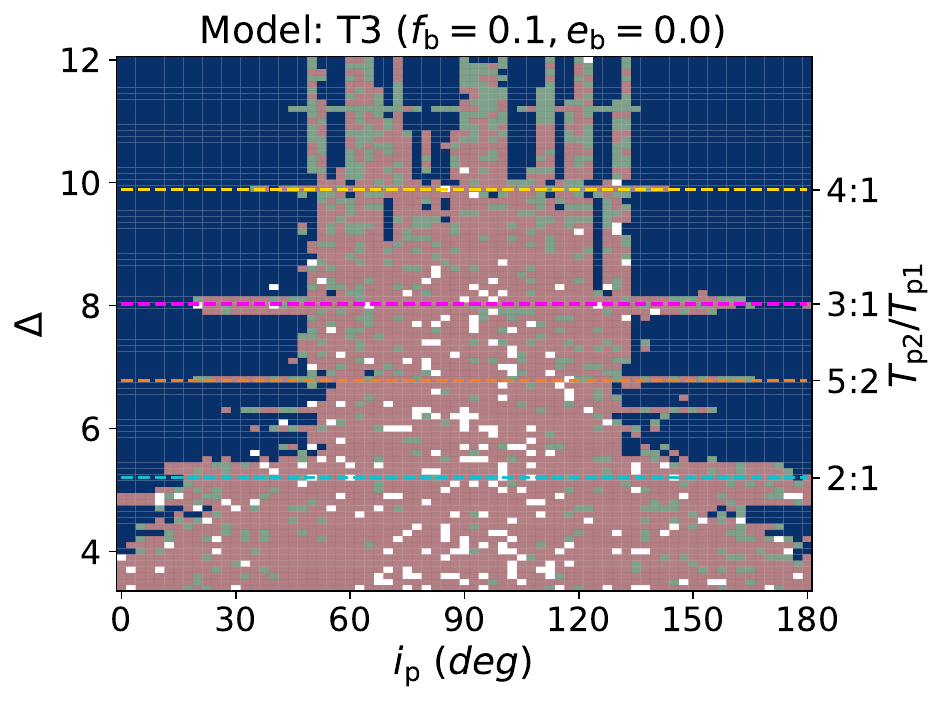}
    \includegraphics[width=8.7cm]{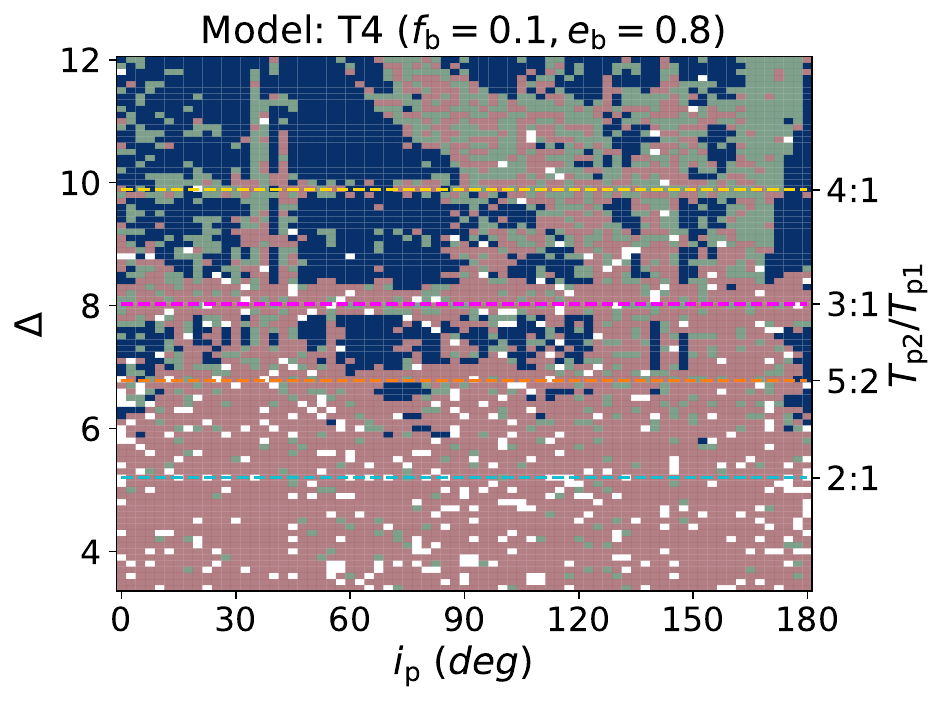}       
    \caption{Same as Fig.~\ref{fig:two} except these are for three planet systems. The blue pixels show stable systems. The green pixels shows systems with 2 stable planets. The red pixels show systems with one stable planet and the white pixels have no stable planets. }
     \label{fig:three}
\end{figure*}

\begin{figure*}
  \centering
    \includegraphics[width=8.7cm]{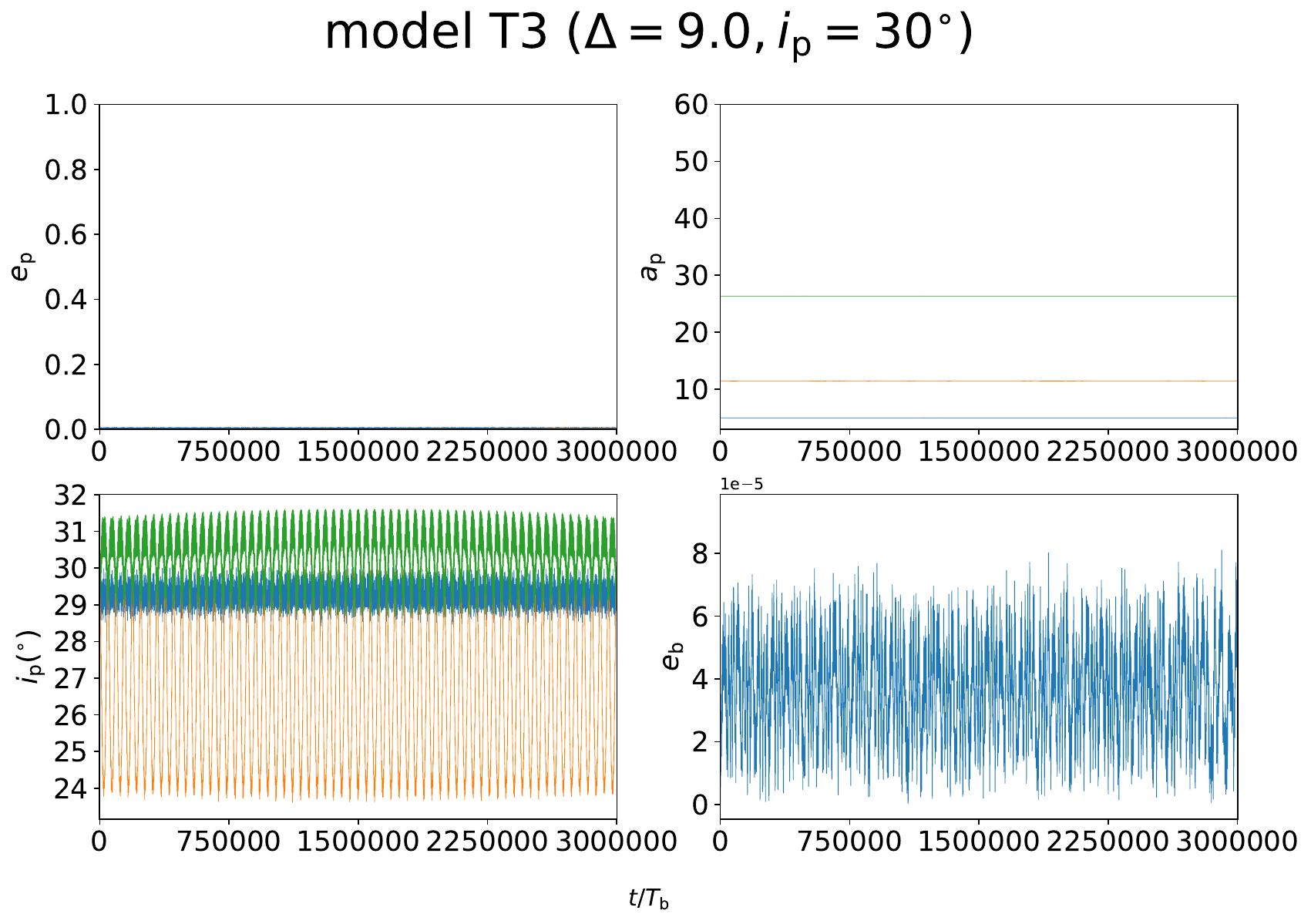}
    \includegraphics[width=8.7cm]{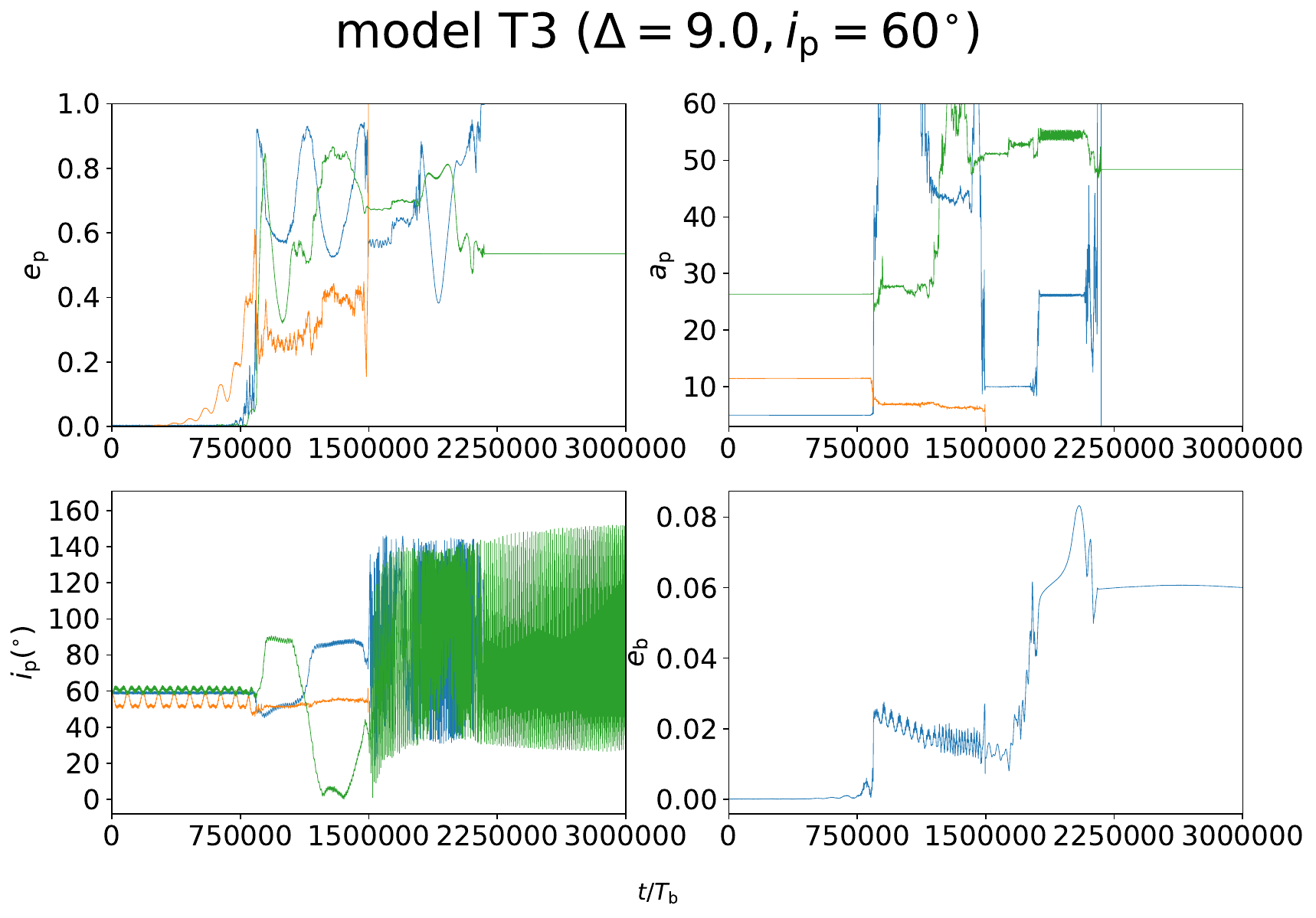}
    \caption{Orbital dynamics in model T3 of Fig.~\ref{fig:three} with $\Delta = 9.0$ and the initial $i_{\rm p} = 30^{\circ}$ (left panels) and $60^{\circ}$ (right panels). Each set of four panels shows the planet eccentricity (upper-left), planet semi-major axis (upper-right), planet inclination (lower-left) and binary eccentricity (lower-right) and includes the inner planet (blue lines), the middle planet (yellow lines) and the outer planet (green lines).}
     \label{fig:T3}
\end{figure*}

Now we consider planetary systems with three CBPs. Fig.~\ref{fig:three} shows the same stability maps as Fig.~\ref{fig:two} except there are now three planets in the system. Since the three planets have equal masses, the semi-major axis ratios of $a_{\rm p2}/a_{\rm p1}$ and $a_{\rm p3}/a_{\rm p2}$ are equal to each other. Consequently, the inner planet pair and the outer planet pair have the same orbital period ratio because $T_{\rm p}\propto	a_{\rm p}^{1.5}$. Therefore, if the inner two planets are in an MMR, so are the outer two  and therefore the three planets are in a Laplace resonance. The four horizontal dashed lines indicate the 2:1, 5:2, 3:1 and 4:1 MMRs of both the inner pair and the outer pair. For convenience, we just label $T_{\rm p2}/T_{\rm p1}$ on the figures.

The stability maps for the three planet cases are qualitatively quite similar to the two planet cases, but with additional instability in all models. Unstable MMR locations become wider in their ranges of both separations and inclinations in the three-planet case than the two-planet case due to the Laplace resonance. Especially around the 2:1 MMR region, three CBPs can only be stable when they are nearly coplanar for model T1.  The vertical unstable belts  are similar to the two planet case but extends to larger $\Delta$. The outcomes of most unstable pixels are single planet survival cases for closely spaced planets with $\Delta \lesssim 8$, while for $\Delta \gtrsim 8$, the most likely outcome is two stable planets.

For the equal mass binary with $e_{\rm b} = 0.8$  (upper-right panel), the regions around the MMRs becomes more unstable with three planets. Fewer systems can be stable within the 2:1 MMR region. Comparing with model T1, the polar region between the 5:2 and 2:1 MMRs are more stable.

For a circular binary with $f_{\rm b} = 0.1$ (lower-left panel), the region between $i_{\rm p}=50^{\circ} \sim 130^{\circ}$ is significantly more unstable compared to the two planet case due to strong ZKL oscillations. In figure 3 of \citet{Chen2023}, we showed that  ZKL oscillations between two CBPs  may cause the instability of the two planet system. In Fig.~\ref{fig:T3} we plot some examples of the evolution of orbital properties of three planets with the time. The four panels shows $e_{\rm p}$ (upper-left), $a_{\rm p}$ (upper-right), $i_{\rm p}$ (lower-left) and $e_{\rm b}$ (lower-right) with time in units of $T_{\rm b}$ for model T3 with $\Delta=9.0$ and $i_{\rm p0}=30^{\circ}$ (left panels) and $60^{\circ}$ (right panels). The blue, yellow and green lines represent the inner, middle, outer planets, respectively.

For the stable case (left) with an initial inclination of $i_{\rm p}=30^{\circ}$, the CBPs undergo tilt oscillations with respect to each other while the inclinations and semi-major axes are nearly constant with time. For the unstable case (right) with initial inclination $i_{\rm p}=60^{\circ}$, the middle and outer CBPs undergo tilt oscillations initially, but the eccentricity $e_{\rm p}$ of the middle planet gets excited through ZKL like behaviour. As a result, the system becomes unstable and the inner two planets are ejected,  only the outer planet survives.

For an eccentric binary with $f_{\rm b} = 0.1$ and $e_{\rm b}$ = 0.8 (lower-right panel of Fig.~\ref{fig:three}), there is significantly more instability compared to the two planet case. There are very few stable cases inside of the 5:2 MMR.  Above there, there are more stable cases but the map shows that three CBPs systems are unlikely to be stable  in the range $i_{\rm p} = 90^{\circ}\sim 170^{\circ}$. Moreover, unstable cases around this region are heavily dominated by two planet surviving cases (green pixels).

\subsection{The number of surviving planets}
\label{dis}

\begin{figure*}
  \centering
    \includegraphics[width=17.4cm]{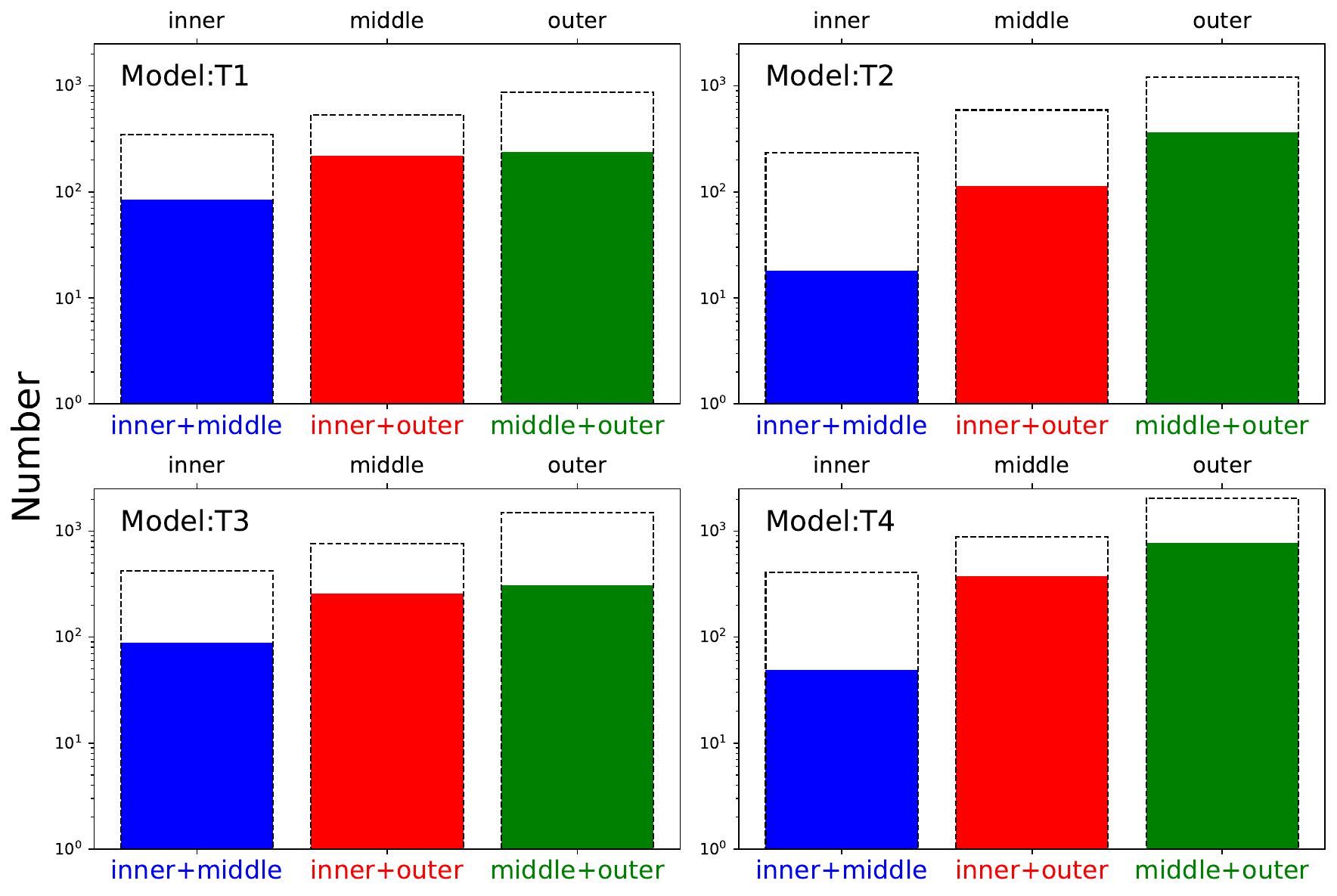}
    \caption{Histograms of the number of surviving planets from the stability maps in Fig.~\ref{fig:three}. The dashed bars show the number of systems with a single surviving planet (the red pixels in Fig.\ref{fig:three}) and the solid bars show the systems with two surviving planets (the green pixels in Fig.~\ref{fig:three}). }
     \label{fig:sh}
\end{figure*}

The complicated orbital interactions of three massive planets and a binary result in different numbers of surviving planets for different binary parameters. At the end of the simulations, the multi-CBP system may be left with three planets, a planet pair, a single CBP or no planets.   Understanding the final distributions of orbital planet properties can help us to predict the multi-CBP system morphology for future discoveries. The abnormal eccentricities of CBPs may be a crucial piece of evidence for planet-planet interactions in the early stages of the system.

The red pixels (single surviving planet cases) in the stability maps dominate most of the unstable outcomes. This suggests that many misaligned CBP systems may only have one surviving massive planet if multiple massive planets form in a compact architecture. In Fig.~\ref{fig:sh}, the dashed bars in each panel show the total number of surviving inner planets, middle planets and the outer planets in  single planet survival cases  in the stability maps of models T1 $\sim$ T4. Because the inner planet is the closest planet to the binary, for model T1, the total survival number of inner planets is lower than the middle and outer planets by factors of 1.5 and 2.5, respectively. With the higher $e_{\rm b}$ = 0.8 of model T2, the factors increase to 2.5 and 5.2. With the lower $f_{\rm b}$ = 0.1 of model T3, factors are 1.8 and 3.6 while with $f_{\rm b}$ = 0.1 and $e_{\rm b}$ = 0.8 of model T4, two factors are 2.2 and 5.0. The similar ratios indicate that there could be abundant CBPs with large distances to their host binaries if multi-circumbinary planetary systems are common in the universe.

The green pixels in the stability maps (two surviving planet cases) dominate most of the unstable orbits above the 3:1 MMR region, $\Delta\gtrsim 8$, in the stability maps. This implies that misaligned CBP systems may have more than one massive CBP if multiple massive planets form in a relatively wide architecture. The solid bars in Fig.~\ref{fig:sh} represent the number of systems with two surviving planets. The bars represent the total surviving numbers of inner and middle planets survived cases (blue), inner and outer planets survived cases (red) and middle and outer planets survived cases (green). The majority of the unstable cases are single planet survival cases so the total number of two planets survived cases is lower than the single planets survival cases. In model T1, the total survival number of inner and outer planets survived cases is lower than the numbers of inner and outer planets survived cases and middle and outer planets survived cases by factors of 2.6 and 2.8. With higher $e_{\rm b}$ = 0.8 of model T2, factors increases to 6.3 and 20.0. With lower $f_{\rm b}$ = 0.1 of model T3, factors are 2.9 and 3.5 while with $f_{\rm b}$ = 0.1 and $e_{\rm b}$ = 0.8 of model T4, two factors are 7.6 and 16.0. 
The inner planet has the highest chance to become unstable so therefore we may find more planet pairs with larger separations in binaries.

\subsection{The eccentricity of the surviving planets}

\begin{figure*}
  \centering
    \includegraphics[width=17.4cm]{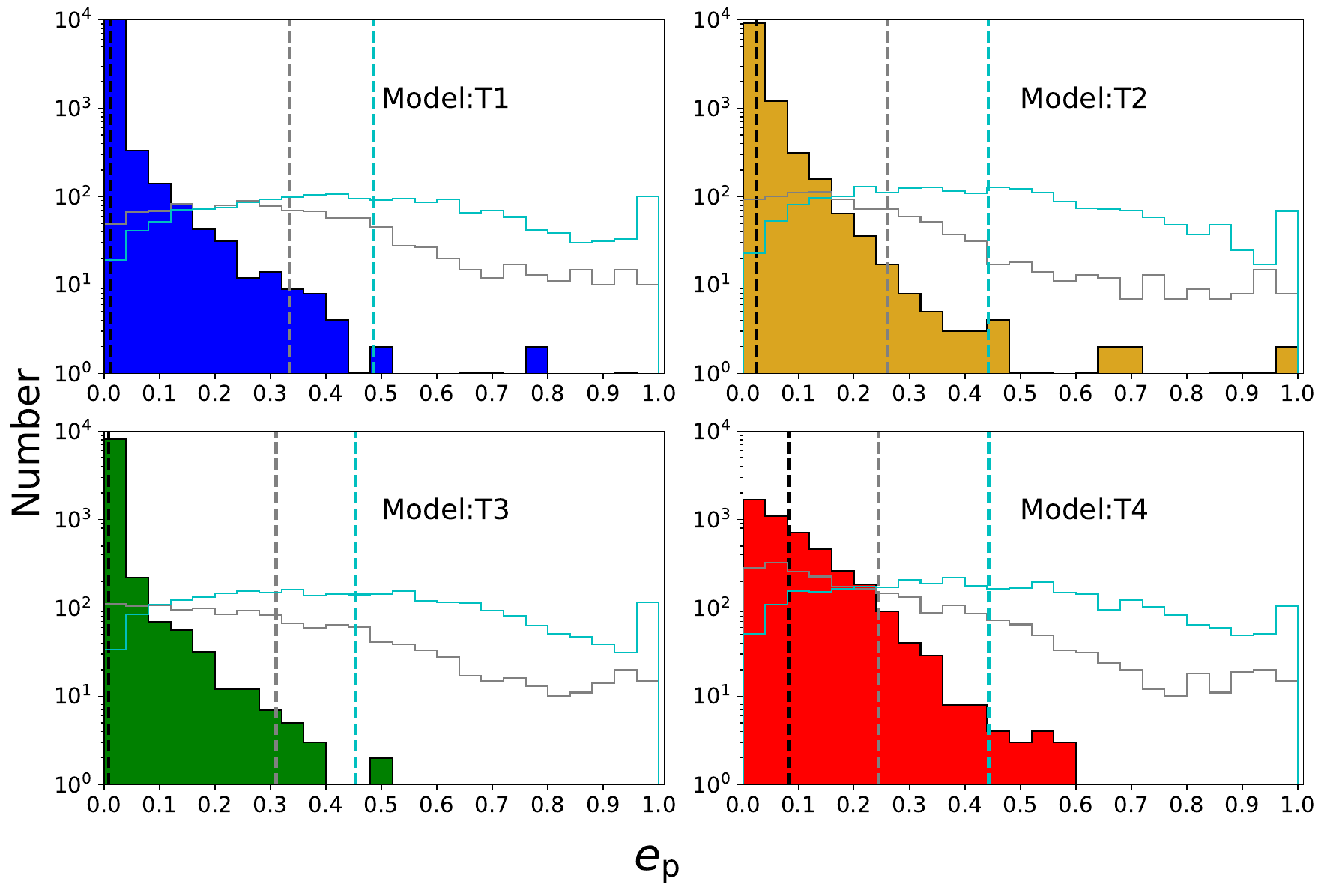}
    \caption{Histograms of the final $e_{\rm p}$ distributions for surviving planets in models T1--T4. The solid histograms show cases in which three planets are stable. The grey  histograms show cases in which two planets are stable and the cyan histograms show cases in which only one planet is stable. The three vertical dashed lines indicate the mean values of each histogram. }
     \label{fig:ee}
\end{figure*}

In this subsection, we now plot histograms of the eccentricity of the surviving planets. In Fig.~\ref{fig:ee}, each histogram represents the sum of the number of inner, middle and outer planets since the differences between the distributions for different planets are small. The solid histograms  are the final eccentricities of planets of models T1--T4 from the blue pixels (three stable planets cases) in the stability maps. The grey empty histograms are those of the green pixels (two stable planets cases)  and the cyan empty histograms are those of from the red pixels (single stable planet cases). The vertical black, grey and cyan lines in each panel represent the mean values of the eccentricities of the solid, grep empty and cyan empty bars, respectively. 

The vertical black lines show that if all three planets are stable, most of planets gain a little eccentricity, even in the most unstable map (model T4). For the vertical grey lines (two stable planet cases), models T1 and T3 with $e_{\rm b} = 0.0$ have larger values ($> 0.3$) than those of with larger $e_{\rm b}$ of models T2 and T4 since the planets in an eccentric binary system have a higher chance to undergo a close encounter with the binary. Thus, planets get ejected before their eccentricities are excited to larger values. Finally, for the vertical cyan lines (single stable planet cases), they have a similar value of about 0.45 in all four panels and these values are larger than the values of two stable planet cases.

\section{Conclusions}
\label{con}

With $n$-body simulations we have modelled a circumbinary planetary system consisting of three giant planets around a circular or eccentric binary. The inner planet is at 5$a_{\rm b}$ initially, where a single planet is stable.  The outer planets are separated by a constant number of mutual Hill radii in the range 3.4--12.0. With this configuration, the period ratio of the inner pair is the same as the outer pair. As a result, if one planet pair is in a MMR, then they are all in a Laplace resonance. 
With stability maps of varying initial planet inclination and separation, we have shown that the binary mass fraction and the binary eccentricity play important roles in the stability of CBPs. Overall,  three CBP systems around an equal mass binary (models T1 and T2) are qualitatively similar to maps of two CBP systems although there is more instability in three planet systems. MMRs are wider in the three planet systems. 
With a low binary mass fraction $f_{\rm b}$ = 0.1 (model T3), there is significantly more instability in a three planet system around $i_{\rm p} = 50^{\circ} \sim 130^{\circ}$ due to strong ZKL oscillations resulting in excitation of planetary eccentricities and ultimately the ejection of planets.

Our simulations show that the single planet survival cases dominate the unstable outcomes for systems with closely spaced planets with $\Delta \lesssim 8$ while  two planet survival cases dominate for more widely spaced systems with $\Delta \gtrsim 8.0$. The surviving planets that were from unstable systems may not be too close to the binary since the inner planet is more likely to be ejected than the other planets. On the other hand, the remaining planets may have significant orbital eccentricities. 

If $\Delta > 10.0$ for exoplanetary systems, MMRs do not play a role in the stability of CBPs. Models T1--T2 show that an equal mass binary can host two or three CBPs with arbitrary inclinations beyond this region. This separation is consistent with the observation that 93$\%$ of planet pairs are at least 10$\,R_{\rm Hill}$ apart \citep{Weiss2018}. A binary system with three compactly spaced massive CBPs is unlikely to remain stable, especially for planets with misaligned orbits to the binary. A system with a relatively wide architecture of three CBPs, separated by more than 8$\,R_{\rm Hill}$, is  likely to remain stable especially for a  close to equal mass binary system.



\section{Implications for exoplanet observations}
\label{implications}

If circumbinary planets are observed to have significant eccentricities, this may imply that the system  had additional planets which were ejected after dynamic interactions with the other planets and the binary. The mean value of the planet eccentricities in cases where two planets survive is slightly smaller than cases in which only a single planet survived. The mean value of the eccentricities of single surviving  planets is similar to the those of with two massive planets initially \citep{Chen2023}. Thus, our results suggest that a single massive CBP which has a significant eccentricity might have been a multi-planetary system initially. However, it is hard to distinguish whether this binary system had hosted two or three massive planets before.
 
Close interactions with the binary result in the inner planet tending to become unstable more easily than the outer planets. The chance of survival of the inner planet is smallest in the single planet survival cases and it decreases with increasing $e_{\rm b}$ and $f_{\rm b}$. The chance of survival of the inner and middle planets in two planets survival cases is also lowest. Currently, half of confirmed CBPs are located close to the edge of stable radii \citep{Yamanaka2019}. However, our study shows that the inner planet that is located at 5$a_{\rm b}$ has a high chance to get ejected if the system has outer planets \citep[see also][]{Chen2023}. Consequently, it may hard to find another massive planet beyond a close-in and misaligned massive planet in those systems unless it is far from the binary. On the other hand, we have not considered different $a_{\rm p1}$ in this study because the outer planet is too far to simulate within reasonable computational time. \citet{Chen2023} showed that different $a_{\rm p1}$ have  different stability maps due to the competition between ZKL oscillations and nodal oscillations. With another distant massive planet outside, we predict their systems will be more unstable if they are misaligned initially.

Multi-CBP systems could be rare due to the intrinsic distribution of planet pairs. The period ratio distribution of Kepler's multi-planet systems indicates that the peak of planet pairs has a period ratio near 2.2. Beyond a period ratio of 2.5, it follows a power law with an exponent -1.26 \citep{Steffen2015}. Our maps show the single CBP survival cases dominate around the 5:2 MMR and thus, the number of giant pairs or triple-giant systems is rare. Nevertheless, this principle may be only valid for the compact multi-planet system. Wide multi-planet single star systems, such as CI Tau, may have 2 inner Jupiter-mass planets and 2 outer Saturn-mass planets ranging from 0.1 to 100 au \citep{Clarke2018} and HR 8799 has four Jupiter-mass planets ranging from 16 to 76 au \citep{Marois2008}.

\begin{figure}
  \centering
    \includegraphics[width=8.7cm]{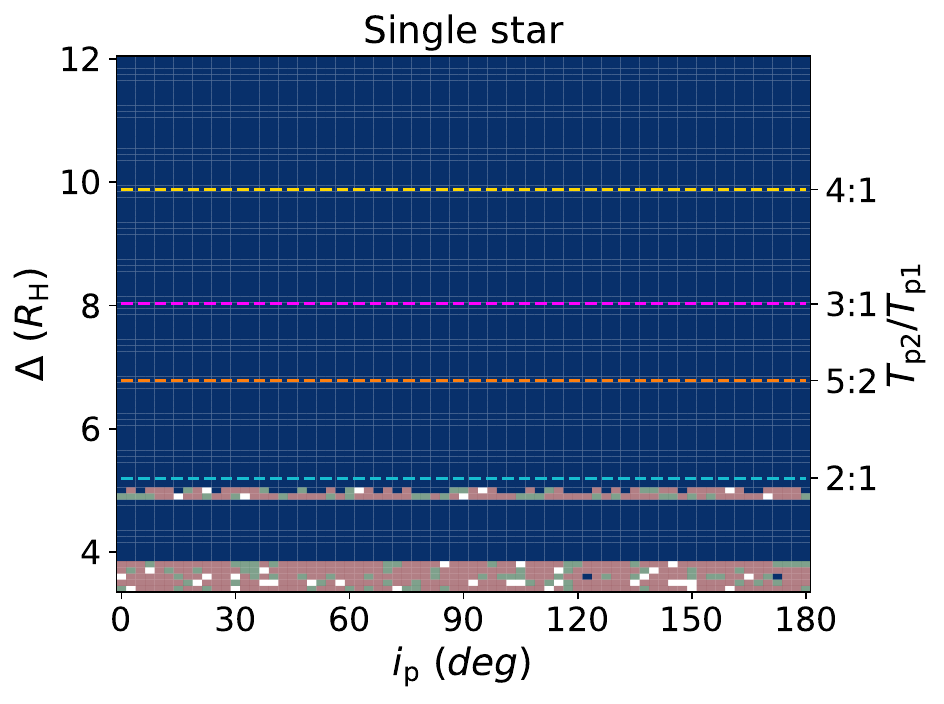}
    \caption{Similar to Fig.~\ref{fig:three} except the centre is a single star and the innermost planet is placed at 5au.}
     \label{fig:single}
\end{figure}

The ejection of planets from a circumbinary system may generate large numbers of free-floating planets.  For comparison, the stability of a system with three Jupiter-mass planets around a single star is shown in Fig. ~\ref{fig:single}. There is only instability around regions of the 2:1 MMR and separation $\Delta < 4.0$. Similarly, for a single star with only two Jupiter-mass planets, only the region with separation $\Delta < 4.0$ is unstable \citep[see Figure 1 in][]{Chen2023}. As a result, a single star system hosting three Jupiter-mass planets is very stable compared to those around a binary. Planet-planet interactions around single stars are unlikely to be a significant contributor to free-floating planets even if there are more than three Jupiter-mass planets unless the system is extremely compact. This conclusion is in agreement with \citet{Veras2012}. In contrast, a multi-CBP system with three massive planets is more unstable than that of with two especially around the MMR regions. Estimates for the mass of free floating planets range from around $0.25$ \citep{Mroz2017} up to about $3.5$ \citep{Sumi2011} Jupiter masses per main-sequence star. Therefore, we suggest that binary systems may contribute most of the free-floating planets due to planet-planet and binary-planet interactions.

This study could contribute to our understanding of circumbinary planet formation and evolution and may help to explain the observational data from the {\it Transiting Exoplanet Survey  Satellite} (TESS) and {\it Planetary Transits and Oscillations of stars} (PLATO). These may find many multiple misaligned circumbinary planets around binaries as a result of the well-developed ETV tools in the near future \citep{Zhang2019}. While we have only considered one relatively close-in initial location for the innermost planet in the system, current observational methods (except gravitational microlensing and direct image observation) tend to detect more close-in CBPs. Therefore, simulations with close-in CBPs are more important. Nevertheless, the ETV method can also detect planets at far distance to the host star, and thus, simulations with multi-CBP system with the large separation to the binary may be necessary for future studies.

Although all confirmed CBPs are nearly coplanar to their binary orbital planes, the observational results from \citet{Czekala2019} have shown the wide distribution of disc inclinations. Comparing with our previous study of two CBPs in \citet{Chen2023}, hosting three stable Jupiter-mass planets around a binary is more difficult than hosting two stable Jupiter-mass planets initially. Besides, if the inner planet is located at 5$a_{\rm b}$, we find that there is less planet pairs can exist the with orbital period ratio < 2:1 even though they satisfy the minimal separation $\Delta \approx 2\sqrt{3}$. A binary with a low mass fraction and a high eccentricity is unlikely to have three Jupiter-mass planets but they can host at least two planets if they are widely separated enough ($\Delta > 10.0$). An equal mass binary has a higher chance to host three  stable Jupiter-mass planets for all initial inclinations since only MMRs between the planet pair contribute the most to destabilise the system. Moreover, the surviving single planets or planet pairs could have significant eccentricities which could be confirmed after more CBPs are found in near future.

\section*{Data Availability}
    The simulations in this paper can be reproduced by using
    the REBOUND code  (Astrophysics Source Code
    Library identifier ascl.net/1110.016). The data underlying 
    this article will be shared on reasonable request to the corresponding author.

\section*{Acknowledgements}
Computer support was provided by UNLV's National Supercomputing Center and DiRAC Data Intensive service at Leicester, operated by the University of Leicester IT Services, which forms part of the STFC DiRAC HPC Facility (www.dirac.ac.uk). CC and CJN acknowledge support from the Science and Technology Facilities Council (grant number ST/Y000544/1). CC thanks for the useful discussion with Dr. Wei Zhu on the PPVII conference. CJN acknowledges support from the Leverhulme Trust (grant number RPG-2021-380). RGM acknowledges support from NASA through grant 80NSSC21K0395. We thank Prof. Yan-Xiang Gong for his careful reading of our manuscript and
gives us many insightful comments and suggestions. Simulations in this paper made use of the REBOUND code which can be downloaded freely at http://github.com/hannorein/rebound.



\bibliographystyle{mnras}
\bibliography{main} 

\bsp
\label{lastpage}
\end{document}